\documentclass[10pt,conference]{IEEEtran}
\IEEEoverridecommandlockouts
 \usepackage{times}
\usepackage{tablefootnote}
\usepackage{balance}

\usepackage[hyphens]{url}
\usepackage{amssymb}
\usepackage{amsmath}
\usepackage{wrapfig}
\usepackage{multirow}
\usepackage{graphicx}
\usepackage{algorithm}
\usepackage{algorithmic}
\usepackage{url}
\usepackage{booktabs}
\usepackage{color}
\usepackage{moreverb}
\usepackage{balance}
\usepackage{amssymb,amsmath}
\usepackage{wrapfig}
\usepackage{multirow}
\usepackage{graphicx}
\usepackage{algorithm}
\usepackage{algorithmic}
\usepackage{epstopdf}
\usepackage{subfigure}
\usepackage{url}

\usepackage{enumitem}
\usepackage{miniplot}
\usepackage{ifthen}

\usepackage{pifont}
\usepackage{bbding}
\usepackage{diagbox}

\usepackage{multirow}
\usepackage{tabularx}
\usepackage{graphicx}
\usepackage{diagbox}
\def\BibTeX{{\rm B\kern-.05em{\sc i\kern-.025em b}\kern-.08em
		T\kern-.1667em\lower.7ex\hbox{E}\kern-.125emX}}

\begin{document}
\title{An Empirical Study of the Landscape of Open Source Projects in Baidu, Alibaba, and Tencent}
	
\author{\IEEEauthorblockN{Junxiao Han\IEEEauthorrefmark{1}, Shuiguang Deng\thanks{*Shuiguang Deng is the corresponding author.}\IEEEauthorrefmark{1}\IEEEauthorrefmark{2}, David Lo\IEEEauthorrefmark{3}, Chen Zhi\IEEEauthorrefmark{1}\IEEEauthorrefmark{2}, Jianwei Yin\IEEEauthorrefmark{1}, Xin Xia\IEEEauthorrefmark{4}}
		\IEEEauthorblockA{\IEEEauthorrefmark{1}College of Computer Science and Technology, Zhejiang University, Hangzhou, China}
		\IEEEauthorblockA{\IEEEauthorrefmark{2}Alibaba-Zhejiang University Joint Institute of Frontier Technologies, Hangzhou, China}
		\IEEEauthorblockA{\IEEEauthorrefmark{3}School of Information Systems, Singapore Management University, Singapore, Singapore}
		\IEEEauthorblockA{\IEEEauthorrefmark{4}Faculty of Information Technology, Monash University, Melbourne, Australia}
		\{junxiaohan, dengsg, zjuzhichen, zjuyjw\}@zju.edu.cn, davidlo@smu.edu.sg, xin.xia@monash.edu}
	
	\maketitle
	
\begin{abstract}
	
Open source software has drawn more and more attention from researchers, developers and companies nowadays. Meanwhile, many Chinese technology companies are embracing open source and choosing to open source their projects. Nevertheless, most previous studies are concentrated on international companies such as Microsoft or Google, while the practical values of open source projects of Chinese technology companies remain unclear. To address this issue, we conduct a mixed-method study to investigate the landscape of projects open sourced by three large Chinese technology companies, namely Baidu, Alibaba, and Tencent (BAT). We study the categories and characteristics of open source projects, the developer's perceptions towards open sourcing effort for these companies, and the internationalization effort of their open source projects. We collected 1,000 open source projects that were open sourced by BAT in GitHub and performed an online survey that received 101 responses from developers of these projects. Some key findings include: 1) BAT prefer to open source frontend development projects, 2) 88\% of the respondents are positive towards open sourcing software projects in their respective companies, 3) 64\% of the respondents reveal that the most common motivations for BAT to open source their projects are the desire to gain fame, expand their influence and gain recruitment advantage, 4) respondents believe that the most common internationalization effort is ``providing an English version of readme files", 5) projects with more internationalization effort (i.e., include an English readme file) are more popular. Our findings provide directions for software engineering researchers and provide practical suggestions to software developers and Chinese technology companies.
\end{abstract}

\maketitle

\section{Introduction}\label{intro}
Over the years, open source software has become one of the cornerstones of modern software development practices \cite{dias2017drives}. Open source movement has dramatically reduced the cost of building and deploying software \cite{coelho2017modern}. Nowadays, more and more organizations and developers rely on open source solutions to sustain and speed up the development of their software projects. Meanwhile, the emergence of modern platforms for developing and maintaining open source projects has boosted the advancement of open source movement \cite{eghbal2016roads,coelho2017modern}. The most notable platform is GitHub; developers use fork operations to create copies of repositories in GitHub, then they improve the copies by submitting a pull request back to the project maintainers \cite{gousios2014exploratory,kalliamvakou2016depth}.

Large software companies like Google, Microsoft, and Facebook have made some of their proprietary software open sources to embrace the open source community. The reasons why they open source proprietary software include fostering innovation, attracting community to engage, and bringing additional benefits to the organization and the community \cite{Facebook,Microsoft_net}. There also exist several studies that investigate the maintenance and evolution of OSS development \cite{lavallee2015good,wu2016maintenance,coelho2017modern,behnamghader2017large}. Kochhar et al. \cite{kochhar2019moving} studied six Microsoft projects to present the transition process from closed to open source. They explore the reasons for open sourcing proprietary software, the steps taken during the transition process, developers' views before and after the transition, etc. 

Recently, more and more Chinese technology companies have participated in and chosen to open source their projects, which has attracted more and more software developers to participate and contribute to. Among them, Baidu, Alibaba, and Tencent (BAT) are the leading and most prestigious technology companies in China, i.e., Baidu has the number one search engine in China, along with many other businesses\footnote{https://content.biostratamarketing.com/blog/an-introduction-to-baidu-china-s-leading-search-engine}, Alibaba owns and operates a diverse array of businesses around the world in numerous sectors, and is named as one of the world's most admired companies by Fortune\footnote{https://www.alizila.com/alibaba-named-fortunes-worlds-admired-companies-list/}\footnote{https://www.investopedia.com/insights/10-companies-owned-alibaba/}, Tencent has been credited as one of the world's most innovative companies by numerous media and firms\footnote{https://www.caixinglobal.com/2018-02-22/tencent-tops-chinese-leaderboard-on-global-innovator-list-101212655.html}\footnote{https://www.chinainternetwatch.com/16343/most-innovative-companies-2015/}\footnote{http://usa.chinadaily.com.cn/epaper/2015-12/03/content\_22617535.htm}.

Despite the aforementioned facts, most studies only investigate activities of software companies like Microsoft \cite{kalliamvakou2015open,devanbu2016belief,kochhar2019moving} or Google \cite{jaspan2018advantages}, while there is no study that perform research on Chinese technology companies, especially BAT. In this paper, we take the first step to perform a research on BAT and study the landscape of projects open sourced by them. To achieve this goal, we conduct a mixed qualitative and quantitative study to investigate the categories and characteristics of projects open sourced by BAT, the developer's perceptions towards open sourcing software projects, and the internationalization effort for such open source projects. Notably, previous studies \cite{kalliamvakou2015open,devanbu2016belief,kochhar2019moving,jaspan2018advantages} performed on Microsoft or Google focused on how commercial organizations use GitHub for collaboration when developing commercial projects, the transition process from closed to open source of proprietary projects, and the advantages and disadvantages of monolithic repositories in large software companies, which are quite different with our study. In our study, we are dedicated to having an overall impression of the characteristics of projects open sourced by BAT.

Consequently, our research can enable software engineering researchers and software practitioners to have a comprehensive and objective understanding of projects open sourced by BAT. It can also provide guidance to improve open source prospects for Chinese technology companies, which, in turn, results in a better open source environment for developers worldwide. The process that we follow in this study is as follows:

\begin{itemize}
	\item To have an intuitive understanding of the existing open source projects, we collected 1,000 open source projects in GitHub that were open sourced by BAT. 
	\item To gain a deeper understanding of the landscape of open source projects (e.g., attitudes, motivations, internationalization, etc.), we sent an online survey to 1,000 developers of these collected open source projects. The survey received 101 responses from the developers.
\end{itemize}

Some of the important findings include: 1) BAT prefer to open source frontend development projects, and most of these projects are frameworks, libraries, and tools, 2) the majority of the respondents (88\%) are positive to open sourcing software projects in their respective companies; however, surprisingly, respondents with the highest experience are the least positive to open sourcing effort, 3) the most common motivations for BAT to open source their projects are ``gain fame, expand influence, and gain recruitment advantage" and ``get feedback from and share to open source community", 4) the most common internationalization efforts are ``providing an English version of readme files" and ``using English in the code comments", 5) projects developed with internationalization effort (i.e., including an English readme file) gain more popularity.

The main contributions of this paper are summarized as follows:
\begin{itemize}
	\item To the best of our knowledge, we are the first to perform a large-scale empirical study to understand the landscape of projects open sourced by BAT.
 	\item Our study combines both qualitative and quantitative analysis; we obtain our findings from an online survey of 101 respondents and a collected dataset of 1,000 open source projects in GitHub that were open sourced by BAT.
 	\item We provide a better understanding of project categories, attitudes towards and motivations of projects open sourced by BAT. Moreover, we also provide a better understanding of internationalization efforts made by these companies to their open source projects.
	\item We highlight some practical implications for Chinese technology companies, practitioners, and researchers. BAT and practitioners should pay more attention to the internationalization effort of their open source projects. We also point to several problems that need more research in the future: e.g., automated tools to support internationalization process. 
\end{itemize}

\section{Research Questions}\label{questions}
In this section, we present the three research questions that we would like to answer in this study.

\vspace{0.2cm}\noindent {\bf RQ1 What are the characteristics of BAT open source projects?}

In this RQ, we want to get an overview of and understanding the types of projects open sourced by BAT.

\vspace{0.1cm}\noindent {\bf RQ2 What are the developers' perceptions towards open sourcing effort at BAT?}

\vspace{0.1cm}\noindent {\bf Attitudes: }The developers' attitudes to open sourcing projects of BAT are likely to affect their enthusiasm to participate in open source projects. We believe that developers with positive attitudes are more likely to be active in contributing to open source projects. Therefore, we want to understand the attitudes of developers and analyze how various demographic factors affect their attitudes.

\vspace{0.1cm}\noindent {\bf Motivations: }Understanding the reasons why BAT open source their projects can provide crucial insights into BAT and their open source projects, which can also highlight ways to promote open source development for other Chinese technology companies.

\vspace{0.2cm}\noindent {\bf RQ3 What factors influence the internationalization effort of projects open sourced by BAT?}

Chinese companies and developers are accustomed to using Chinese in their projects. Although such behaviors are beneficial to speed up the development process, it is very detrimental to the internationalization effort of the projects after open sourcing. Therefore, we want to investigate into internationalization effort done by BAT, and analyze the impact of such effort. 

\section{Methodology}\label{methodology}
\subsection{Overview}

To answer the three research questions, we follow a mixed-method approach, including qualitative and quantitative analysis. The data are collected from two sources: (1) open source projects in GitHub that were open sourced by BAT, and (2) responses to our survey that we sent to 1,000 developers of the collected open source projects. The collected open source projects and the survey and responses are publicly available at \url{https://github.com/hanjunxiao/BAT-projects-survey}. 

\subsection{Project Analysis}\label{Online_Data}

\subsubsection{Data Collection}\label{Data_Collection}

We obtained all open source projects in GitHub under all accounts (official account and team accounts) of BAT, to make the analysis more comprehensive. Notably, since our goal is to understand the landscape of projects open sourced by BAT, we did not differentiate open source projects whether they were internal projects that got open-sourced or they were new projects that were created open source. As a result, we obtained 702 projects from Baidu, 1,173 projects from Alibaba, and 133 projects from Tencent, respectively. Consequently, we obtained 2,008 projects in total. All the data were collected up to April 5th, 2019.

Then, we removed projects that were forked or empty, removed projects with less than 6 commits (following \cite{kalliamvakou2016depth}) and 2 contributors (following \cite{kalliamvakou2016depth,han2019characterization}). Ultimately, 1,000 open source projects remained, which are analyzed in this paper. The statistics of the projects are shown in Table \ref{statistics_projects}. We extracted the information of open source projects via GitHub API. 

\begin{table}[t]
	\centering
	\scriptsize
	\caption{The statistics of the dataset for BAT open source projects in GitHub.}
	\vspace{-0.3cm}\begin{tabular}{p{1.5cm}<{\centering}|p{1.5cm}<{\centering}|p{1.5cm}<{\centering}|p{1.5cm}<{\centering}}
		\hline
		\textbf{Company} & \textbf{Number of accounts in BAT} & \textbf{BAT projects in GitHub} & \textbf{Studied projects}\\
		\hline
		Baidu & 13 & 702 & 380\\
		\hline
		Alibaba & 16 & 1,173 & 520\\
		\hline
		Tencent & 4 & 133 & 100\\
		\hline
		Total & 33 & 2,008 & 1,000\\
		\hline
	\end{tabular}%
	\label{statistics_projects}\vspace{-0.5cm}%
\end{table}%

\subsubsection{Categorization}\label{Categorization_Characteristics}

To observe what types of projects BAT have open sourced, we performed card sorting \cite{spencer2009card} to categorize these collected projects by manually looking into the full name, description and readme content. 

\vspace{0.1cm}\noindent\textbf{Step 1: Card Sorting.} We first selected the open source projects under the official account (baidu, alibaba, tencent) of BAT and manually observed the project name, descriptions, and readme contents. Then, we manually clustered the open source projects into different groups with respect to their functionality. Lastly, we discussed each group and determined its name by referring to the categories defined in Sharma et al.'s study \cite{sharma2017cataloging} and the SourceForge website\footnote{https://sourceforge.net/directory/os:windows/}. The first and fourth authors collaboratively determined the categories of open source projects. As a result, we obtained 9 categories, which can be seen in Table \ref{category}. Notably, since some projects cannot be fitted to the 9 categories, we thus put these projects in the ``Other" category and this category will not be discussed in the rest of the paper.

\begin{table*}[t]	
	\centering
	\scriptsize
	\caption{Classification Categories.}\label{category}
	\vspace{-0.3cm}\begin{tabular}{|p{1.7cm}<{\centering}|p{5.2cm}<{\centering}|p{1.8cm}<{\centering}|p{7.7cm}<{\centering}|}
		\hline
		\textbf{Category} & \textbf{Description} & \textbf{Abbreviation} & \textbf{Examples} \\
		\hline
		Backend Development & Projects related to server-side development, typically corresponding to tools and libraries. & BD & Project name: Tencent/mars; Description: Mars is a cross-platform network component developed by WeChat; URL: https://github.com/Tencent/mars\\
		\hline
		Frontend Development & Projects related to user interface and client-side development, typically corresponding to tools and libraries. & FD & Project name: Tencent/weui; Description: A UI library by WeChat official design team, includes the most useful widgets/modules in mobile web applications; URL: https://github.com/Tencent/weui\\
		\hline
		Machine Learning & Projects that provide machine learning tools, datasets, models, etc. & ML & Project name: alibaba/x-deeplearning; Description: An industrial deep learning framework for high-dimension sparse data; URL: https://github.com/alibaba/x-deeplearning\\ 
		\hline
		Management and Monitoring & Projects that develop solutions for system and code management or monitoring. & MM & Project name: Tencent/OOMDetector; Description: OOMDetector is a memory monitoring component for iOS which provides you with OOM monitoring, memory allocation monitoring, memory leak detection and other functions; URL: https://github.com/Tencent/OOMDetector\\
		\hline
		Operating System & Projects that develop operating systems or are related to operating systems & OS & Project name: alibaba/cloud-kernel; Description: Alibaba Cloud Linux Kernel - an open-source Linux kernel originated by Alibaba Operating System Team; URL: https://github.com/alibaba/cloud-kernel\\
		\hline
		Program Analysis & Projects that develop tools to analyze source code statically or dynamically in development or production environment, such as parsers, compilers, testing and debugging tools, etc. & PA & Project name: alibaba/arthas; Description: Alibaba Java Diagnostic Tool; URL: https://github.com/alibaba/arthas\\
		\hline
		Storage System & Projects that develop storage systems, such as new databases, file systems, key-value storage systems, etc. & SS & Project name: Tencent/MMKV; Description: An efficient, small mobile key-value storage framework developed by WeChat. Works on iOS, Android, macOS and Windows; URL: https://github.com/Tencent/MMKV\\
		\hline
		Web Service/ Cloud Computing & Projects used to manage web services and cloud computing resources, such as service agents, cluster managers, etc. & WC & Project name: alibaba/nacos; Description: an easy-to-use dynamic service discovery, configuration and service management platform for building cloud native applications; URL: https://github.com/alibaba/nacos\\
		\hline
		Webpage & Projects that are static websites only used for displaying or sharing information. & Wp & Project name: weixin/WeIndex; Description: Index for WeChat-related resources; URL: https://github.com/weixin/WeIndex\\
		\hline
		Other & Projects that cannot fit to any one of the above categories, such as games, text editors, etc. & Oe & Project name: AlloyTeam/StreetFighter; Description: A game open sourced by AlloyTeam; URL: https://github.com/AlloyTeam/StreetFighter\\
		\hline
	\end{tabular}\vspace{-0.4cm}
\end{table*}%

\vspace{0.1cm}\noindent\textbf{Step 2: Labeling.} Next, the first and fourth authors independently labeled all the open source projects under all accounts of BAT. After that, Fleiss Kappa \cite{fleiss1971measuring} is computed to measure the agreement between the two labelers. The Kappa value between two labelers is 0.89, where an almost perfect agreement is reached. After the manual labeling process, the two labelers reach a common decision by eliminating their disagreements. For projects with ambiguous classification categories, the two labelers discuss to determine its final classification categories.

\subsection{Survey}\label{Survey}

Some of the research questions that we want to investigate cannot be answered by just analyzing the projects from GitHub. Thus, to complement the GitHub data, we designed an online survey to help us better answer the research questions, particularly related to the attitudes and motivations to open source projects, and the internationalization effort. Survey was anonymous with the intention to increase response rate \cite{tyagi1989effects}.

\subsubsection{Survey Design}\label{Survey_Design}

\begin{table*}
	\scriptsize
	\centering
	\caption{Survey Questions}\label{survey_content}
	\vspace{-0.3cm}\begin{tabular}{p{0.8cm}<{\centering}p{0.8cm}<{\centering}p{9cm}<{\centering}p{6cm}<{\centering}}
		\hline
		\textbf{RQ*} & \textbf{QN} & \textbf{Question Description} & \textbf{Answer Options} \\
		\hline
		D & Q1 & Do you work as a company employee (including intern, outsourcing employees) or a volunteer contributor for the open source projects in GitHub? & [Company employee / Volunteer contributor] \\
	
		D & Q2 & Which of the following roles best describe your software engineering experience? & [Frontend Development / Backend Development / Software Testing / Project Management / Other] \\
		
		&   & With how many years of experience?  & [\#] / [\#] / [\#] / [\#] / [\#] \\
		
		D & Q3 & What is the primary programming language used in your development process? & [\#] \\
		
		D & Q4 & Where is your current country of residence? & [\#] \\
		
		RQ2 & Q5 & What is your attitude for open source software practices of Chinese technology companies? & [Positive / Neutral / Negative] \\
		
		
		&   & \emph{Q6 only appeared based on the answer to Q5} &  \\
		
		RQ2 & Q6 & Please describe the reasons for selecting `Negative': & [\#] \\
		
		RQ2 & Q7 & What is the motivation and expected impact for open sourcing software projects in companies? & [\#] \\
		
		RQ3 & Q8 & Have you considered internationalization in your current projects? & [Yes / No] \\
		
		&   & \emph{Q9 and Q10 only appeared based on the answer to Q8} &  \\
		
		RQ3 & Q9 & Please describe your efforts for improving internationalization if your choice is `Yes': & [\#] \\
		
		RQ3 & Q10 & Please give some reasons if your choice is `No': & [\#] \\
		\hline
	\end{tabular}

\emph{*numbers refer to the research questions corresponding to the survey questions, 'D' refers to demographic questions.}\vspace{-0.5cm}
\end{table*}

Table \ref{survey_content} lists the survey questions (note that some questions are open-ended), the research questions corresponding to the survey questions, and the provided answer options. We collect respondents' demographic information including whether they are company employees or volunteer contributors, their job roles and work experiences (in years), primary programming language, and current country of residence. Thus, we can filter out the respondents who may not have a deep understanding of our survey and make it convenient to segment our survey results by demographic groups. We also asked respondents about the attitudes to, motivations of, and their internationalization effort for projects open sourced by BAT. In this way, we can answer the research questions better.

\subsubsection{Participant Selection}\label{Participant_Selection}
We recruited respondents with the following strategy: we first obtained the email addresses publicly listed as owners to the open source projects of BAT in GitHub. As a result, 14 available email addresses were obtained. After that, we obtained the email addresses publicly listed as contributors to the open source projects of BAT in GitHub. By removing the empty and duplicate email addresses, 6,344 unique email addresses were obtained. Therefore, we got 6,358 email addresses in total and randomly chose 1,000 email addresses to send our survey to avoid spamming. 

\subsubsection{Pilot Survey}\label{Pilot_Survey}
To ensure the quality and clarity of our survey, we conducted a pilot survey with respondents who are Computer Science professors and graduate students with experience in open source practices. Based on their feedback, we made some minor edits. Responses from the pilot survey were excluded from our analysis. Due to that the majority of the respondents being Chinese, we prepared the survey in Chinese.

\subsubsection{Data Analysis}\label{Data_Analysis}
After a period of 25 days, we received 101 responses, which resulted in a response rate of 10.1\% (101/1,000). For free-form answers, we applied an open card sort \cite{spencer2009card} to cluster them. The card sort was composed of two phases: we first created a card for each question response. Then, we assigned the cards into meaningful groups, where each group accompanied with a descriptive title. No groups were predefined in this process, where we let the groups come out through this process. 

\section{Results}\label{result}
In this section, we present the results to the project analysis and survey questions to answer the three research questions proposed in Section \ref{questions}.

\subsection{Overall Impressions}\label{Overall_Impressions}

\subsubsection {Survey Results}

Our research is conducted to study the complex landscape of projects open sourced by BAT, where the majority of the participants are Chinese. Thus, not surprisingly, most of the respondents reside in China, and only 5 respondents reside abroad, i.e., Brazil, Germany, Japan, Kenya, and Russia.

We then excluded one response from the respondents whose job role is neither frontend development, backend development, testing, nor project management. The job role of the excluded respondent is described as technical management. At the end of the filtering step, we obtained 100 responses, which we will further analyze.

It can be observed that 37 of the respondents are company employees, which accounts for 37\% of the total respondents. The rest of the respondents are volunteer contributors, which takes up a large proportion of the overall respondents.

Moreover, the number of years of work experience of the respondents ranged from 0.25 to 20 years, where the average number of years is 4.5 years. Out of all the respondents, 64\% described their job role as frontend development, 74\% described their job role as backend development, 9\% described their job role as software testing, and 17\% described their job role as project management. It is worth noting that since some respondents may take multiple roles, the percentages do not add up to 100\%. 

\subsection{RQ1: What are the characteristics of BAT open source projects?} \label{result_rq1}
The findings for this research question come from our manual analysis and card sorting described in Section \ref{Categorization_Characteristics}. We first investigate the distribution of the 1,000 open source projects into the 9 categories shown in Table \ref{category}. Then, we compare the popularity of open source projects that belong to different categories. Lastly, we provide discussions including implications for Chinese technology companies and open source developers.

\subsubsection{Category Distribution}

\begin{figure}[bt]
	\centering
	\includegraphics[width=1\linewidth]{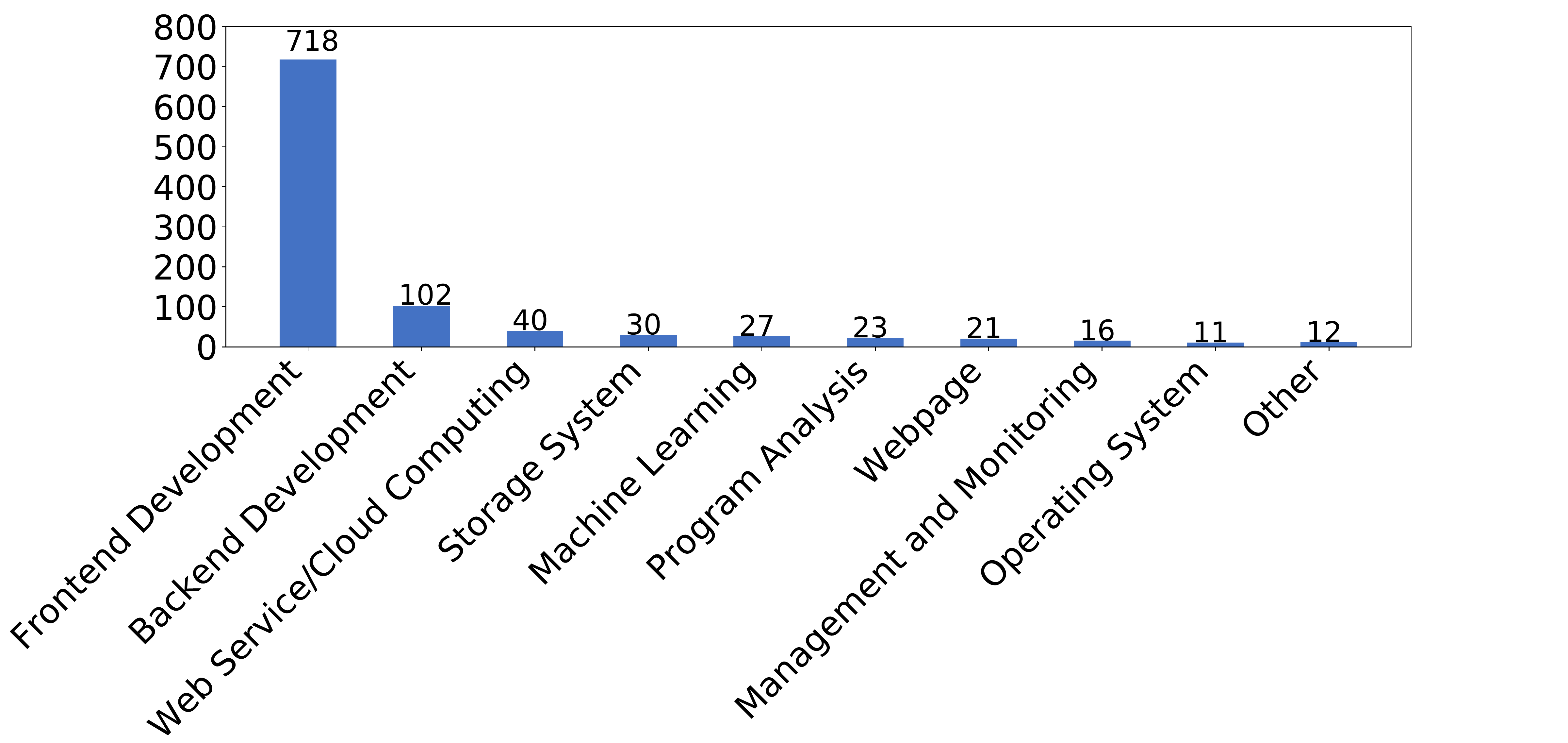}
	\vspace{-0.7cm}\caption{Distribution of Categories.}\vspace{-0.5cm}
	\label{rq2_BAT_category_distribution}
\end{figure}

Fig. \ref{rq2_BAT_category_distribution} shows the distribution of different categories. We can conclude that the dominant category of projects that were open sourced by BAT is frontend development, which accounts for 71.8\% of the total. The second most common category is backend development, which takes up 10.2\% of the total. Although the backend development category accounts for the second highest number of occurrences, the number of such projects is far less than the frontend development category. Conversely, operating system and management and monitoring categories have the least occurrence in our studied projects, which only account for 1.1\% and 1.6\% of the total, respectively.

The findings suggest that BAT prefer to open source frontend development projects. According to our manual observation, we find that most of these frontend development projects are frameworks, libraries, and tools. For example, Alibaba open sourced the \emph{ice} project, which is a simple and friendly frontend development framework, they also open sourced the \emph{ant-design} project, which is a UI design language and React UI library.

\subsubsection{Category Popularity}

\begin{figure}[t]
	\centering
	\includegraphics[width=0.8\linewidth]{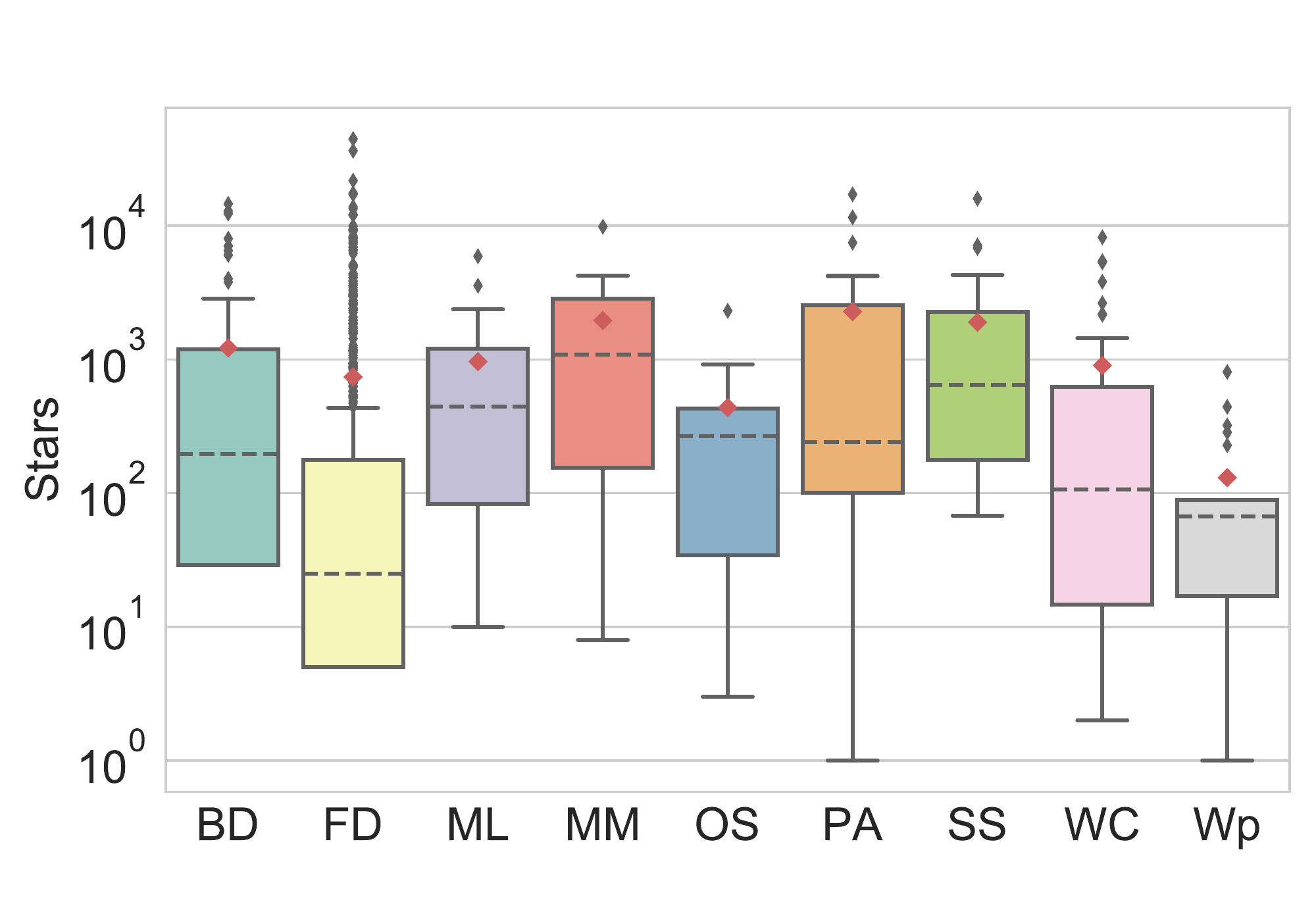}
	\vspace{-0.5cm}	\caption{Distribution of stars of different categories of studied projects.}
	\label{rq2_star}\vspace{-0.6cm}
\end{figure}

Here, we use the number of stars as the popularity metric \cite{pinto2018challenges,han2019characterization} to describe the distribution of popularity of different categories, which can be seen in Fig. \ref{rq2_star}. It reveals that the largest number of stars is for a project in the frontend development category (44,436) named \emph{ant-design}\footnote{\url{https://github.com/ant-design/ant-design}}. We also find that the smallest number of stars for a frontend development project is 0, and the frontend development category has the smallest median in terms of the number of stars (25), which indicates that projects with this category present a widespread in popularity and most projects with this category have a low popularity.

To explore why projects with the frontend development category exhibit a widespread in popularity, we manually analyzed the frontend development projects. Consequently, we find that some companies choose to open source a frontend development framework with many modules to provide a flexible combinable solution. The main framework and optional modules are open sourced as separate projects, so that users can use the main project (the main framework) and many other optional projects (the optional modules) according to their needs. For example, Baidu open sourced the saber (361 stars) project - a frontend mobile framework aiming to provide a solution for mobile web development. They also open sourced many of its modules as separate projects to provide different functions, e.g., saber-cookie (3 stars), saber-ui (1 star), saber-widget (2 stars), saber-scroll (9 stars), saber-dom (6 stars), saber-firework (7 stars), etc. We can find that the projects with low popularity are modules, and users tend only to star the main framework but not all modules, which may give one reasonable explanation about this phenomenon.

When it comes to the median measure, the management and monitoring category has the maximum median value of stars (1085.5) -- the project with the highest number of stars in this category is {\em RAP}\footnote{\url{https://github.com/thx/RAP}}. In terms of the mean measure, the category with the highest mean value of stars is program analysis (2,279) -- the project with the highest number of stars in this category is {\em fastjson}\footnote{\url{https://github.com/alibaba/fastjson}}. Then, we applied Wilcoxon Signed-Rank test \cite{wilcoxon1992individual,fan2018chaff} at a 95\% significance level with a Bonferroni correction \cite{abdi2007bonferroni} to investigate whether the popularity of different categories are statistically significantly different. We also computed Cliff's delta \cite{cliff1993dominance} to show the effect sizes. Results show that the mean for program analysis is statistically significantly larger than those of frontend development, web service/cloud computing, and webpage categories, and the effect sizes are at least small. The means for management and monitoring and storage system are statistically significantly larger than those of backend development, frontend development, operating system, web service/cloud computing, and webpage categories, and the effect sizes are also at least small.

\subsubsection{Discussion}

Results in RQ1 reveal that the mean values of stars for projects with management and monitoring and storage system categories are statistically significantly larger than those projects with other categories. Meanwhile, BAT prefer to open source frontend development projects, while open source projects with frontend development category show lower popularity. Therefore, companies and developers should pay more attention to improving the popularity and quality of projects under the frontend development category. They may achieve it by comparing with projects under management and monitoring and storage system categories, to find out why projects under these two categories have a higher popularity. 

\subsection{RQ2: What are the developers' perceptions towards open sourcing effort at BAT?}\label{result_rq2}
The findings for this research question come from survey questions Q5 to Q8 (see Table \ref{survey_content}). We first show the percentage of practitioners whose attitudes are positive/neutral/negative to open sourcing projects in BAT. Then, we present the motivations for BAT to open source their projects from the perspective of project developers. Lastly, we provide discussions including practical implications for Chinese technology companies, open source developers, and researchers.

\subsubsection{Attitudes}

\begin{figure}[t]
	\centering
	\includegraphics[width=1\linewidth]{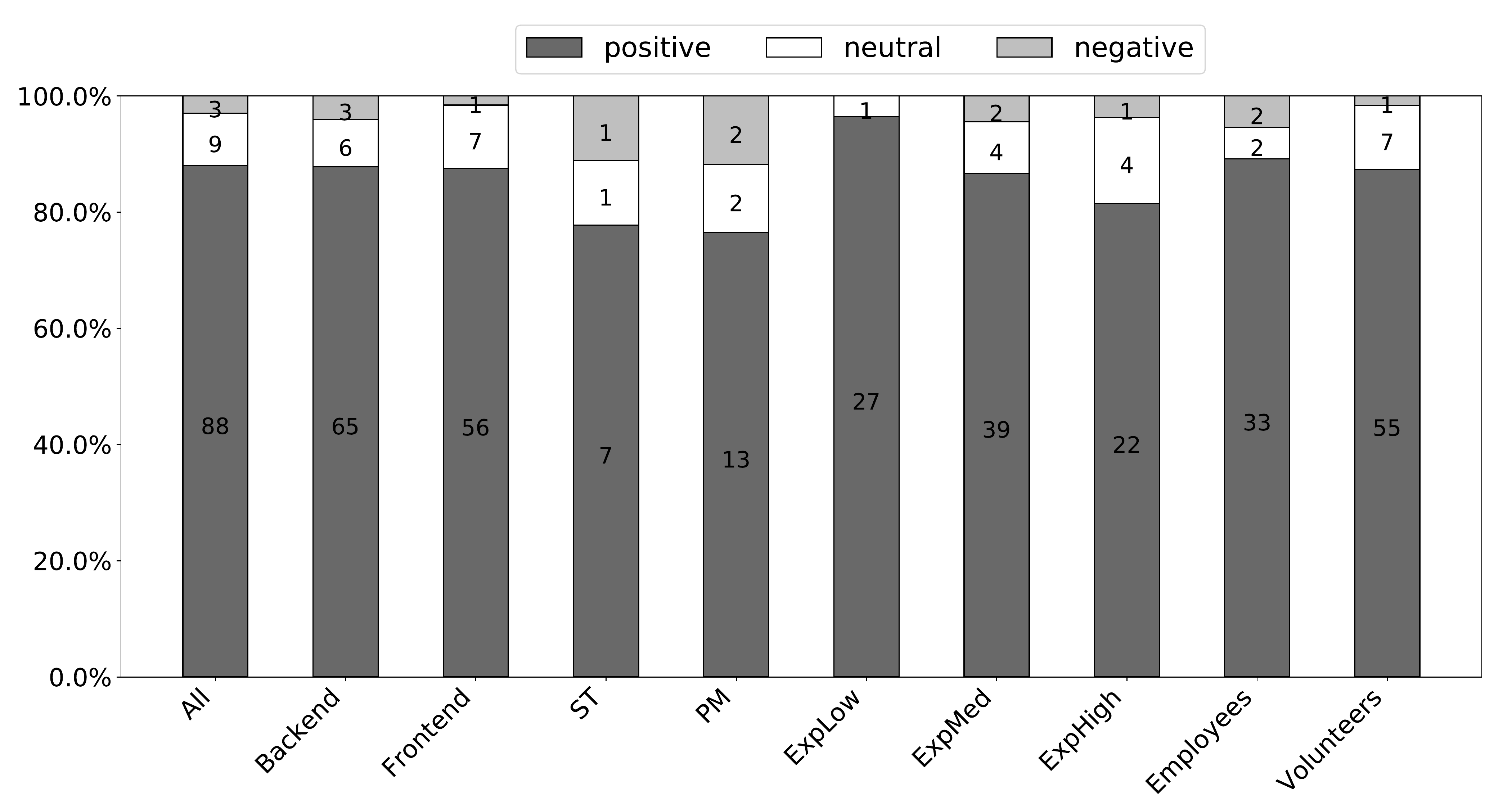}
	\vspace{-0.5cm}	\caption{Percentages of respondents whose attitudes are positive/neutral/negative to open sourcing BAT projects across different demographic groups.}
	\label{rq1_attitude}\vspace{-0.5cm}
\end{figure}

Fig. \ref{rq1_attitude} shows the percentage of practitioners whose attitudes are positive/neutral/negative to open sourcing BAT software projects across different demographic groups. The following demographic groups are considered:

\begin{itemize}
	\item All respondents (All)
	\item Respondents who are company employees/volunteer contributors (Employees/Volunteers)
	\item Respondents who are frontend/backend developers (Frontend/Backend), software testers (ST), or project managers (PM)
	\item Respondents whose experience is low, medium, or high, which are developers whose experience (in years) are in the bottom 25\%, 25\%--75\%, or top 25\% (ExpLow/ExpMed/ExpHigh) 
\end{itemize}

Fig. \ref{rq1_attitude} shows that across all demographic groups, the majority of respondents are positive to open sourcing software projects in BAT, which account for 88\% of the total. Conversely, only 3\% of the respondents are negative to open sourcing efforts in BAT. The remaining 9\% of the respondents are neutral. We also run Wilcoxon Signed-Rank test \cite{wilcoxon1992individual} with a Bonferroni correction to investigate whether the differences of the demographics are statistically significant. We also computed Cliff's delta \cite{cliff1993dominance}. We can observe several differences between the demographics (all are statistically significant at a p-value $<$ 0.05 with an effect size that is at least medium):

\begin{itemize}
	\item Frontend developers and backend developers are statistically more positive to open sourcing effort in BAT than software testers and program managers.	
	\item Respondents with high experience are statistically less positive to open sourcing effort; as experience decreased, participants are more positive to open sourcing effort. 	
\end{itemize}
	
Besides, results show that there is no statistically significant difference between employees and volunteers as for the attitudes of open sourcing effort.	
	
\subsubsection{Motivations}

Next, we present the motivations that prompt companies to open source their software projects from developers' perspective. These are derived from answers to survey question Q7 analyzed using an open card sort process (as described in Section \ref{Survey_Design}). We discuss each motivation, provide the number of respondents that mention each of them, introduce the number of employees and volunteers that support each motivation, and give examples of answer fragments associated with each of them. As each response may contain more than one motivation, the sum of the supporting respondents across all motivations is more than 100.

\noindent\textbf{M1. Gain fame, expand influence, and gain recruitment advantage:} The most common motivation for companies to open source their software projects is the desire to gain fame, expand influence and gain recruitment advantage over competitors. This view is supported by 64 respondents, including 24 employees (65\% of this group) and 40 volunteers (63\% of this group). Similar views were expressed by Kochhar et al. \cite{kochhar2019moving}, i.e., open sourcing projects can help companies build trust with users and developers. Besides, Kochhar et al. \cite{kochhar2019moving} and a blog \cite{reasonstoopen} all believe that open sourcing can improve recruitment and retention, as the blog says, ``having an open source presence attracts and retains top talent." Some of the answers that support this view are given below:

\begin{itemize}
	\item[\ding{47}] \emph{Enhance brand awareness and master the right to speak in the open source field.} (U68)
	\item[\ding{47}] \emph{[\ldots] gain fame and reduce recruitment costs.} (U88)
	\item[\ding{47}] \emph{Establish a brand, expand company influence, and benefit to recruitment.} (U80)
\end{itemize}

\noindent\textbf{M2. Get feedback from and share to open source community:} According to 28 respondents, by open sourcing their internal projects, BAT would like to get feedback from and share what they have to open source community. Among them, 9 employees (24\% of this group) and 19 volunteers (30\% of this group) support this motivation. This result is also in line with the findings of Kochhar et al. \cite{kochhar2019moving} to some extent. They found that open sourcing code helps to get feedback from the community faster. Some survey responses corresponding to this motivation category are:

\begin{itemize}
	\item[\ding{47}] \emph{Take it from open source and give back to open source.} (U12)
	\item[\ding{47}] \emph{Share, then get feedback and improvement from the community.} (U94)
	\item[\ding{47}] \emph{Solve the pain points in industries and propose new solutions.} (U61)
\end{itemize}

\noindent\textbf{M3. Attract more people to use, participate to improve the quality and maturity of projects:} According to 26 respondents, they believe that the companies open source their software projects to attract more people to use and participate, which in turn, improves the quality and maturity of projects. Among the respondents, 7 employees (19\% of this group) and 19 volunteers (30\% of this group) support this motivation. Some examples of responses that support this category are:

\begin{itemize}
	\item[\ding{47}] \emph{Attract more people to use, find and fix bugs faster.} (U81)
	\item[\ding{47}] \emph{Enhance the vitality of the community, hope to attract more developers to join us and actively contribute to the community.} (U31)
	\item[\ding{47}] \emph{Use their own experience to facilitate the development and implementation of related technologies.} (U37)
\end{itemize}

\noindent\textbf{M4. Construct their software ecosystem:} According to 8 respondents, one of the motivations to open source is to build a software ecosystem -- these respondents include 3 employees (8\% of this group) and 5 volunteers (8\% of this group). To illustrate this, consider the Dubbo project, which is a microservice architecture open sourced by Alibaba. Alibaba also released the Nacos project, which is an easy-to-use dynamic service discovery, configuration and service management platform for building cloud native applications. Dubbo and Nacos as well as a series of other open source projects that are open sourced by Alibaba (e.g., Sentinel) formed the shared service system, thus constituted the Dubbo-centered open-source software ecosystem. Some sample responses are:

\begin{itemize}
	\item[\ding{47}] \emph{Build a software ecosystem for chip.} (U59)
	\item[\ding{47}] \emph{Expand the software ecosystem.} (U69)
\end{itemize}

\noindent\textbf{M5. Promote commercial projects:} According to 6 respondents, they consider that one of the motivations to open source is to promote commercial projects -- these 6 respondents include 3 employees (8\% of this group) and 3 volunteers (5\% of this group). Notably, some companies are apt to open source a simplified version of their internal commercial projects. And if users want more features when using the open-sourced simplified version, they will have to pay for the commercial version to obtain full features. In this way, companies can get additional revenues. Some survey responses that support this category are:

\begin{itemize}
	\item[\ding{47}] \emph{Most domestic companies [\ldots] generate business opportunities for internal commercial projects which are commercial versions based on open source projects. } (U44) 
\end{itemize}

\noindent\textbf{M6. Help developers in BAT become better coders/testers:} According to 2 respondents, they think that the companies open source their projects with the intention to help developers in BAT become better coders/testers indirectly. The 2 respondents encompass 1 employee (3\% of this group) and 1 volunteer (2\% of this group). This corroborates previous studies \cite{kochhar2019moving} that mention that open-sourcing code can help internal developers do things that they might have never thought before and help the community become more vigorous. A sample response that supports this category is:

\begin{itemize}
	\item[\ding{47}] \emph{It indirectly requires employees to write high-quality code.} (U55)
\end{itemize}

\noindent\textbf{M7. Other:} There also exist 5 respondents whose responses do not fit any of the above 6 categories, are unclear, or are irrelevant to BAT's motivation of open sourcing projects.

\subsubsection{Discussion}

The motivation M5 reveals that some companies open source a simplified version of their commercial projects. If users want more features when using the open-sourced simplified version, they will be induced to pay for the commercial version to obtain full features. In this way, companies can get additional business profits. This finding suggests that more efforts are needed to balance the relationship between open source effort and profitability for companies. Researchers can investigate this kind of open source projects and their corresponding commercial projects in-depth to find win-win solutions and best practices that companies can adopt to contribute well to the open source community, while at the same time not hurting their bottom line.

\subsection{RQ3: What factors influence the internationalization effort of projects open sourced by BAT?}\label{result_rq3}
Findings for this research question come from survey questions Q8 to Q10 (see Table \ref{survey_content}). We first show the percentage of practitioners who have considered internationalization in their projects. Then, we elaborate the efforts that practitioners made for improving internationalization of their open source projects and introduce why some respondents do not consider internationalization. After that, we make an analysis of the 1,000 collected open source projects. Lastly, we provide discussions including implications for Chinese technology companies, open source developers, and researchers.

\subsubsection {Survey Results}

Question Q8 of our survey asks whether respondents have considered internationalization in the projects. Results show that most of them do, as 73\% of the respondents replied positively. This indicates that internationalization is considered by BAT.

\begin{figure}[t]
	\centering
	\includegraphics[width=1\linewidth]{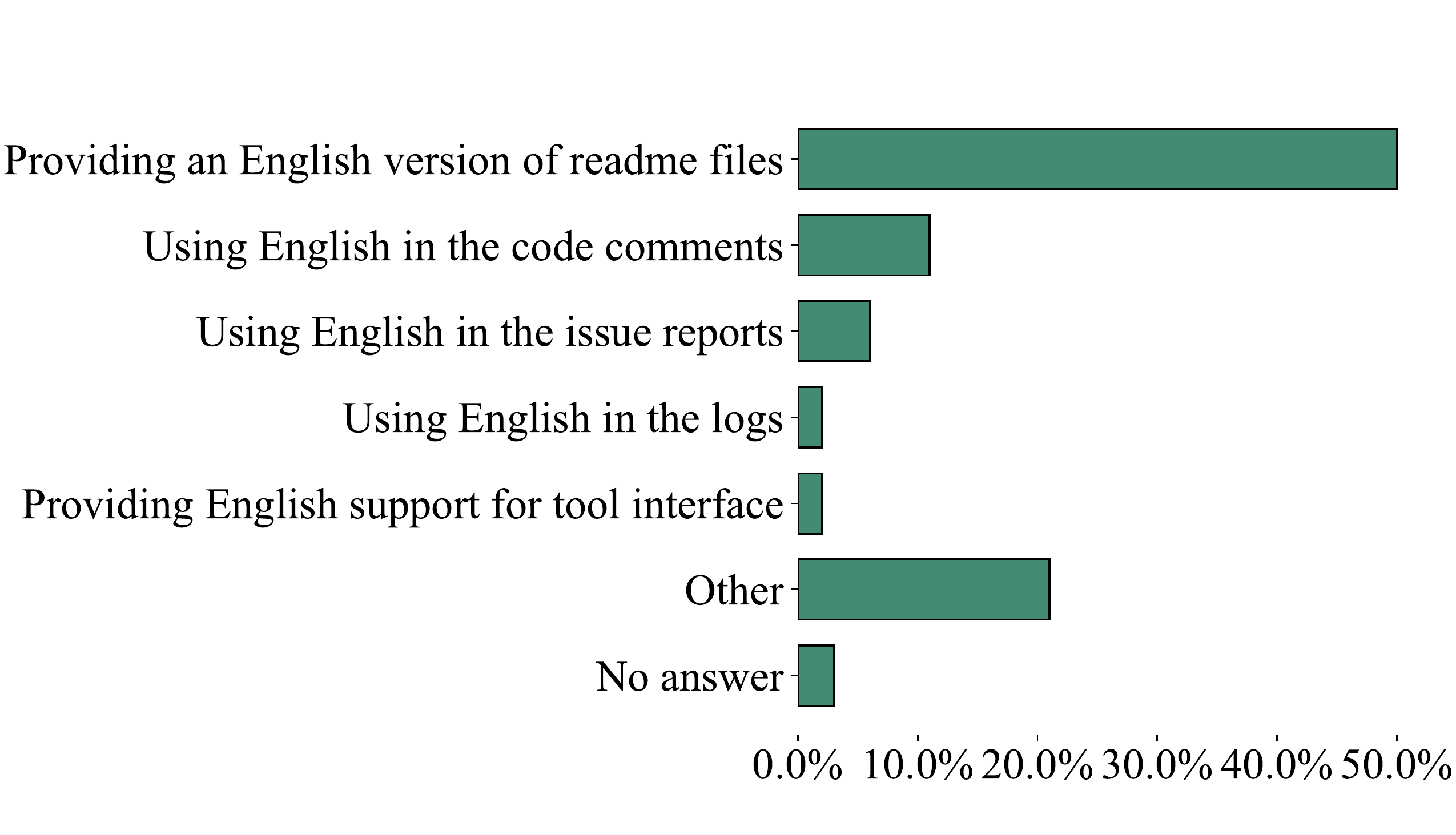}
	\vspace{-0.5cm}\caption{Efforts for internationalization in BAT open source projects.}
	\label{rq3_internationalization_effort}\vspace{-0.5cm}
\end{figure}

\noindent\textbf{Efforts for internationalization. }If respondents say `Yes' to survey question Q8, they will have the opportunity to describe their efforts for improving internationalization of open source projects by providing open-ended comments to Q9. Fig. \ref{rq3_internationalization_effort} illustrates the distribution of answers to Q9 after card sorting. We can observe that the most common effort for internationalization is providing an English version of readme files, followed by using English in the code comments and using English in the issue reports. Moreover, the least common efforts for internationalization are using English in the logs and providing English support for tool interface. The remaining responses in ``Other" and ``No answer" groups are ambiguous or incomprehensible, thus we do not discuss them.

\noindent\textbf{Reasons for no internationalization effort.} For respondents who say `No' to survey question Q8, they will have the opportunity to elaborate the reasons by providing free-form answers in Q10. As a result, the most mentioned reason is limited time and energy (11 respondents), which indicates that these respondents have no extra time to consider internationalization issues. The second most mentioned reason is no need for internationalization (7 respondents), which implies that their open source projects have no intention to enter the international market. Besides, another reason is the size of projects or the number of users is small (3 respondents), which all give some explanations for why projects are not internationalized. Apart from the three reasons, we also have 6 respondents giving other ambiguous responses, and we do not discuss it.

\subsubsection{Project Analysis}  
We then analyze the 1,000 open source projects to compare the popularity between projects with more internationalization effort and those with less (or even no) internationalization effort. As depicted in Fig. \ref{rq3_internationalization_effort}, most of our survey respondents mentioned that one of the most common internationalization efforts is to provide an English version for readme files. Therefore, in this analysis, we use {\em availability of a readme file written in English} as an indicator to determine internationalization effort. That is, we put projects into the ``High" (internationalization effort) group if they have an English readme file, and ``Low" (internationalization effort) group otherwise. In this way, we classify 552 projects into the ``High" group and 448 projects into the ``Low" group.

\begin{figure}[t]
	\centering
	\includegraphics[width=0.8\linewidth]{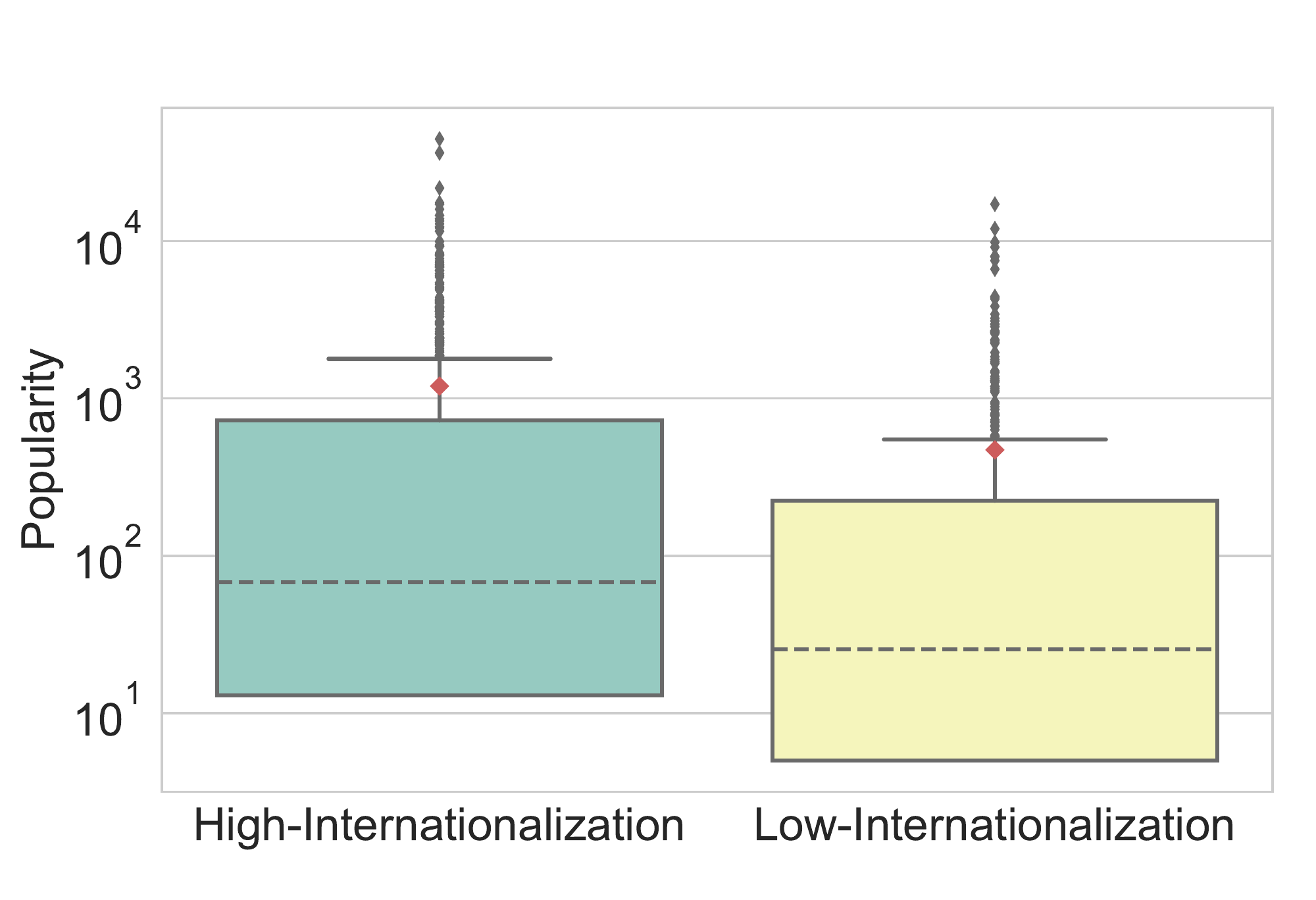}
	\vspace{-0.5cm}\caption{Comparison of popularity of different groups of studied projects.}
	\label{rq3_stars}\vspace{-0.5cm}
\end{figure}

Then, we present the distribution of popularity (the number of stars) of the two groups, which can be seen in Fig. \ref{rq3_stars}. We can find that projects with more internationalization effort received more stars (1,197 vs 471, mean) than those projects with less internationalization effort. To examine whether the difference of the two groups is statistically significant, we apply the Wilcoxon rank-sum test \cite{mann1947test,fan2018chaff} at a 95\% significance level. Then, we use Cliff's delta \cite{cliff1993dominance} to show the effect size of the difference. Results show that there indeed exists a statistically significant difference for popularity between the two groups, and the effect size is small.

\begin{figure}[t]
	\centering
	\includegraphics[width=1\linewidth]{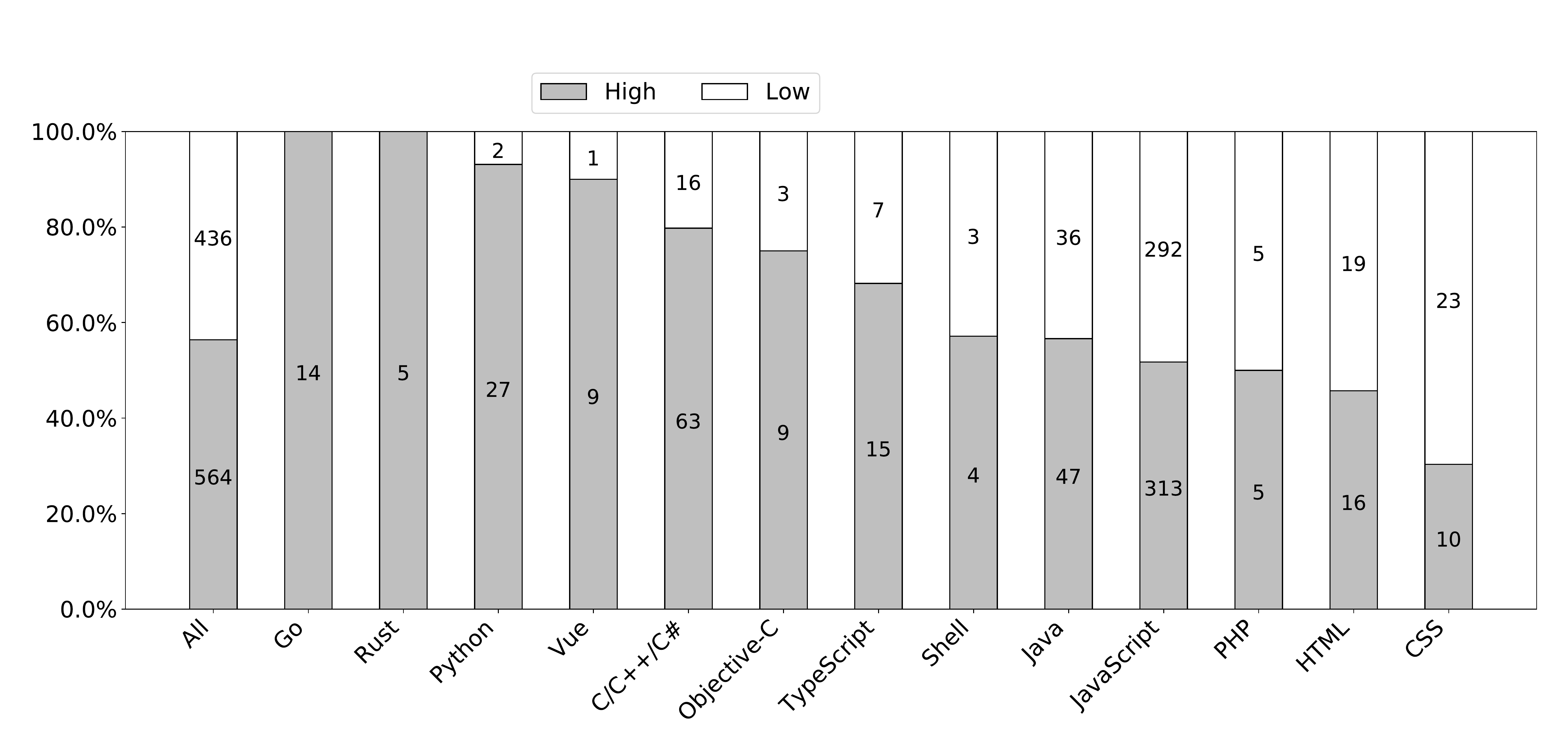}
	\vspace{-0.5cm}\caption{Percentage of projects whose internationalization effort are high/low across various programming languages.}
	\label{rq3_internationalization_language}\vspace{-0.4cm}
\end{figure}

Moreover, we present the percentage of open source projects whose internationalization effort are high/low across various programming languages, which is illustrated in Fig. \ref{rq3_internationalization_language}. Note that programming languages that are only used by one or two projects are omitted in the figures. We can notice that projects with Go, Rust, Python, and Vue programming languages are the top four groups that have high internationalization effort. Projects with CSS, HTML, and PHP programming languages are the bottom three groups. To examine whether the differences of internationalization effort across different programming languages are statistically significant, we perform a Fisher's exact test \cite{fisher1922interpretation,wan2018perceptions} with the null hypothesis - the projects with different programming languages equally have high internationalization effort. The result shows no statistically significant difference.

\begin{figure}[t]
	\centering
	\includegraphics[width=1\linewidth]{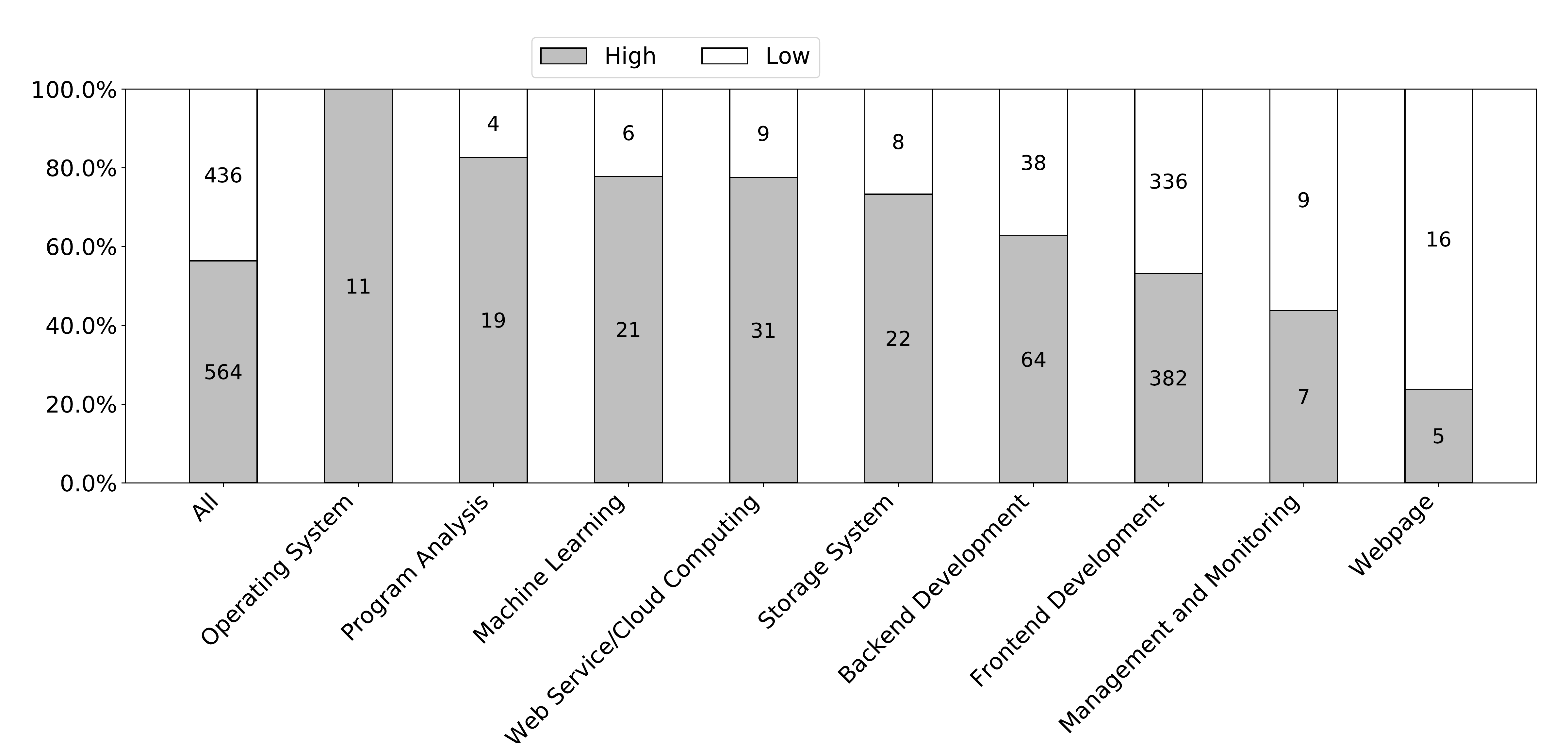}
		\vspace{-0.5cm}\caption{Percentage of projects whose internationalization effort are high/low across different categories.}
	\label{rq3_internationalization_category}\vspace{-0.6cm}
\end{figure}

Notably, projects written in CSS, HTML, and PHP all belong to the frontend development category. Therefore, we also investigate the percentage of open source projects whose internationalization effort are high/low across different categories, which can be seen in Fig. \ref{rq3_internationalization_category}. We can observe that projects belonging to operating system, program analysis, and machine learning categories are the top three groups that have high internationalization effort. Projects belonging to webpage, management and monitoring, and frontend development categories are the bottom three groups. To examine whether the differences of internationalization effort across different categories are statistically significant, we perform a Fisher's exact test again with the null hypothesis - the projects across different categories equally have high internationalization effort. Results show no statistically significant difference.

\subsubsection{Discussion}

Results in RQ3 show that 73\% of the respondents said that they had considered internationalization of open source projects in our survey, while our project analysis shows that only 552 out of 1,000 (55.2\%) open source projects have considered internationalization (i.e., including English readme files). It implies that companies and developers need to do more things to improve open source projects' internationalization process in practice. Meanwhile, our results show that open source projects with high internationalization effort show higher popularity than projects with low internationalization effort. It also highlights the value that Chinese technology companies can reap to pursue internationalization.

We also recommend companies to establish an internationalization standard when they open source software projects. In this way, projects open sourced by Chinese technology companies can be used not only by Chinese developers, but also by developers worldwide. Therefore, it can facilitate the development of the open source community and provide more quality projects to developers worldwide. To achieve that, developers should pay more attention to the internationalization effort of open source projects and form good internationalization-friendly habits, such as providing an English version for readme files, using English in the code comments, using English in the issue reports and pull requests, etc.

Moreover, there also exist ample opportunities for researchers to study. They can give companies guidance on forming an internationalization standard from a research perspective so that companies can establish an internationalization standard more systematically. They can also plow deeply about the internationalization process to highlight several pain points and design automated or semi-automated tools to help reduce the time spent in the internationalization effort.

\section{Threats To Validity}\label{threats}
\textbf{Threats to Internal Validity.} One threat relates to the clarity of questions in our survey. To mitigate this threat, we have done a pilot study, and refine the survey based on inputs that we received from pilot study participants.

Another threat is related to translation work that we need to do. Most of the practitioners we surveyed are Chinese and their native language is Chinese. Thus we prepared our survey in Chinese. The first and fourth authors translated our respondents' answers. Some information may be lost or incorrectly conveyed in this translation process. To mitigate this threat, other authors provided comments to our translations. We also refined the translated text several times to improve its quality.

Another potential threat relates to the response rate. We sent our survey to 1,000 contributors, and received 101 responses, corresponding to a response rate of 10.1\%. Although the response rate is not very high, it is similar to those of past studies that survey software developers \cite{lee2017understanding,latoza2006maintaining,lo2015practitioners}. 

\noindent \textbf{Threats to External Validity.} The results in our paper may not generalize to all Chinese technology companies. In this paper, we analyze the open source projects of BAT in GitHub, and only send our survey to contributors and owners of the projects that are open sourced by BAT. Since we only analyze large companies, results of this paper may not translate to small and medium-sized companies. This threat is mitigated as BAT developers are likely to have worked for other companies before they join BAT (attrition in IT jobs is high). Moreover, our survey respondents include volunteers.

\section{Related Work}\label{related}
\noindent {\bf Studies on Industry-Backed Software Projects in GitHub. }Kalliamvakou et al. \cite{kalliamvakou2015open} performed an online survey and interview with GitHub users to study how GitHub is used for collaboration in commercial projects. Pinto et al. \cite{pinto2018challenges} conducted an exploratory study on eight proprietary projects that became open source to identify some challenges when open-sourcing proprietary projects. They found that only a few projects experienced a growth in newcomers, contributions, and popularity after open-sourcing. Kochhar et al. \cite{kochhar2019moving} studied six Microsoft projects to present the transition process from closed to open source. They conducted an interview and survey to address why large organization open source proprietary software, understand the steps during the transition process, explore the transition's outcomes and challenges, and the OSS community's response to the open-sourcing of projects, etc.

Pinto et al. \cite{pinto2018challenges} and Kochhar et al.'s \cite{kochhar2019moving} studies are the most closely to ours. However, there are differences between their works and ours. Instead of studying the challenges when open-sourcing proprietary projects in Pinto et al.'s work or the transition process from closed to open source of proprietary projects in Kochhar et al.'s work, we focus on the landscape of projects that have already been open sourced by large Chinese technology companies - BAT. We apply project analysis and survey to concentrate on characteristics of such open source projects, explore developer's attitudes to and motivations of open sourcing effort as well as the internationalization effort. Kochhar et al. analyze one research question: why large organizations open source their proprietary software. Our study also investigates this question; the findings of our study confirm some of their findings: open sourcing the code can help companies build trust with users and developers, help them find and hire potential employees, help to get feedback from the community faster, help developers in the company write better code and become better coders/testers. However, our study also provides some insights that have not been discovered in their work, i.e., companies open source projects to construct the company's software ecosystem.

\noindent {\bf Studies on Open Source Software (OSS). }Due to the rapid development of OSS, there is a considerable body of work about it. Foucault et al. \cite{foucault2015impact} studied the impact of developer turnover on quality in open source software. They found that the activities of external newcomers can have a negative impact on software quality. Kazman et al. \cite{kazman2015evaluating} presented a case study of a widely used OSS project to evaluate the impacts of architectural documentation. Singh et al. \cite{singh2017entropy} used a non-homogeneous poisson process model to understand the fixing of issues across releases of open source software. Candido et al. \cite{candido2017test} studied the usage and impact of test suite parallelization in open source projects and found that only 24\% of projects include test suites. Different from the aforementioned studies, we focus on the landscape of projects open sourced by BAT.

\section{Conclusion and Future Work}\label{conclusion}

This paper reports findings of our exploratory study of the landscape of open source projects in BAT. We perform a mixed-method study to gain insights into characteristics of BAT open source projects (RQ1), developers' perceptions towards open sourcing effort at BAT (RQ2), as well as the internationalization effort of BAT open source projects (RQ3). We conduct our analysis by collecting 1,000 projects from GitHub that were open sourced by BAT and conducting a survey to developers contributing to these projects. Our results suggest that 88\% of respondents are positive towards open sourcing software projects, respondents with high experience are more negative to open sourcing, while respondents with low experience are more positive. There is also room for future research to dig in-depth on why participants with different demographics show different attitudes towards open sourcing effort, why Chinese companies prefer to open source frontend development projects, etc.

In the future, we also plan to include more Chinese companies of different sizes (and also potentially other companies from other less studied regions) to investigate the generalizability of our findings.

\section*{Acknowledgment}
{This research was partially supported by the National Science Foundation of China (No. U20A20173, No. 61772461) and Natural Science Foundation of Zhejiang Province (No. LR18F020003).}

\bibliographystyle{IEEETrans}
\bibliography{references}

\end{document}